\documentclass[apj]{emulateapj}
\usepackage{apjfonts}
\usepackage{amsmath}

\usepackage{booktabs}

\begin{document}


\title{Enhanced Dissipation Rate of Magnetic Field in Striped Pulsar Winds by the Effect of Turbulence}

\author{Makoto Takamoto}
\affil{Department of Physics, Kyoto University, Oiwake-cho, Kitashirakawa, Sakyo-ku, Kyoto 606-8502, Japan}
\email{takamoto@tap.scphys.kyoto-u.ac.jp}

\author{Tsuyoshi Inoue}
\affil{Department of Physics and Mathematics, Aoyama Gakuin University, Fuchinobe, Chuou-ku, Sagamihara 252-5258, Japan}
\email{inouety@phys.aoyama.ac.jp}

\author{Shu-ichiro Inutsuka}
\affil{Department of Physics, Graduate School of Science, Nagoya University, Furo-cho, Chikusa-ku, Nagoya 464-8602, Japan}
\email{inutsuka@nagoya-u.jp}




\begin{abstract}
In this letter 
we report on turbulent acceleration of the dissipation of magnetic field 
in the postshock region of a Poynting flux-dominated flow, 
such as the Crab pulsar wind nebula. 
We have performed two-dimensional resistive relativistic magnetohydrodynamics simulations of subsonic turbulence 
driven by the Richtmyer-Meshkov instability at the shock fronts of the Poynting flux-dominated flows in pulsar winds.
We find that turbulence stretches current sheets 
which substantially enhances the dissipation of magnetic field, 
and that most of the initial magnetic field energy is dissipated within a few eddy-turnover times.
We also develop a simple analytical model for turbulent dissipation of magnetic field that agrees well with our simulations.
The analytical model indicates that the dissipation rate does not depend on resistivity 
even in the small resistivity limit. 
Our findings can possibly alleviate the $\sigma$-problem in the Crab pulsar wind nebulae.
\end{abstract}


\keywords{magnetic fields, magnetohydrodynamics (MHD), relativistic processes, shock waves, plasmas, turbulence}



\section{\label{sec:sec1}Introduction}
Magnetic fields play an important role 
in various astrophysical phenomena. 
In particular, 
Poynting flux-dominated plasmas have been studied extensively 
as origins of high energy astrophysical phenomena, e.g., ultra relativistic jets~\citep{1997ApJ...482L..29M}, 
gamma ray bursts (GRB)~\citep{2003astro.ph.12347L}, 
and pulsar winds~\citep{1984ApJ...283..694K,1984ApJ...283..710K}. 
The key point of these models is that 
most of the energy is transferred as Poynting flux from the central engine first, 
and then converted into the thermal energy later.
This makes it easy to form high Lorentz factor conditions and well collimated jets 
since the inertia of the plasma is negligible. 
However, 
it is widely known that 
it is very difficult to sufficiently dissipate the electromagnetic energy 
by simple collisional Ohmic dissipation 
within the observationally indicated characteristic times of phenomena, 
so that many alternative mechanisms have been proposed~\citep{2003NewAR..47..667M,2006A&A...450..887G}. 
Analogous problems can also be found in non-relativistic phenomena, 
such as solar flares~\citep{1984SoPh...94..341S}, 
and it seems to be a generic problem in high magnetic Reynolds number media. 
Therefore, 
research into mechanisms of efficient dissipation of electromagnetic energy is extensive. 

The $\sigma$-problem is a puzzle of energetics between the Crab pulsar wind and pulsar wind nebula (PWN)~\citep{1984ApJ...283..694K,1984ApJ...283..710K}. 
Theoretical studies of the Crab pulsar magnetosphere suggest that 
the pulsar wind is a Poynting flux-dominated flow with $\sigma \simeq 10^{4}$, 
where $\sigma$ is the ratio of the Poynting energy flux to the particle energy flux. 
Furthermore, 
if one assumes an axisymmetric stationary relativistic magnetohydrodynamic (RRMHD) flow, 
the plasma continues to be Poynting flux-dominated far beyond the fast magnetosonic point. 
On the other hand, 
observations of the expansion speed and luminosity of the Crab PWN indicate that 
the plasma should be particle energy-dominated near the inner edge of the nebula as $\sigma \simeq 3 \times 10^{-3}$~\citep{1984ApJ...283..694K,1984ApJ...283..710K}. 
To resolve this problem, 
efficient dissipation processes are necessary in the wind region. 
One of the well-known models is the ``striped wind model''~\citep{1971CoASP...3...80M}
in which oblique rotating pulsars have many radially outgoing  current sheets around the equatorial plane in the wind region.
This model was expected to resolve the $\sigma$-problem through dissipation in these current sheets. 
It was, however, shown that 
the magnetic field energy cannot be fully dissipated 
and Poynting energy still dominates at the inner edge of the Crab PWN~\citep{2001ApJ...547..437L,2003ApJ...591..366K}. 
Concerning this problem, 
P\'etri \& Lyubarsky (2007) and Sironi \& Spitkovsky (2011) 
have performed particle-in-cell (PIC) simulations 
that shown the magnetic field dissipation rate is substantially enhanced by the interaction of a weak precursor MHD shock wave and a current sheet 
that induces a driven magnetic reconnection. 
These works largely advanced our understanding of the $\sigma$-problem, 
which focused mainly on kinetic effects. 
In this letter, 
we show that hydrodynamical effects can also provide an additional mechanism of magnetic field dissipation behind the precursor MHD shock wave. 
The preshock region of the pulsar wind can be inhomogeneous due to many possible sources of fluctuation, 
such as magnetic reconnection at the Y-point close to the light cylinder; 
the fluctuations in the preshock flow can trigger instabilities and generate turbulence in the postshock region, 
such as the Richtmyer-Meshkov instability (RMI) driven by the shock-density fluctuation interaction
~(Giacalone \& Jokipii 2007; Sironi \& Goodman 2007; Inoue et al. 2009,2010,2011,2012). 
In such a turbulent medium, 
current sheets in the striped wind can be stretched by turbulent vortical motions, 
which can enhance the dissipation of magnetic energy.
In the following, 
we consider decaying turbulence by driving it only in the initial setup, 
as in the case of shock-induced turbulence by the RMI 
in which it is only induced at the shock front. 
Such mechanism of magnetic field dissipation is analogous to the turbulent reconnection proposed 
by Lazarian \& Vishniac 1999
(see also Kowal et al. 2009). 
While turbulent reconnection assumes externally driven turbulence, 
our results show that the fast dissipation of magnetic fields proceeds even without external forcing, 
which strengthens the applicability of our mechanism. 

\section{\label{sec:sec1a}Energetics to Drive Turbulence}
In the aforementioned scenario, 
the turbulence induced in the postshock region is driven by thermal and magnetic energy released in the high-$\sigma$ shock.
Hence, the velocity dispersion of the turbulence is expected to be less than the slow magnetosonic velocity behind the high-$\sigma$ shock wave.
In this section, 
we estimate the upper limit of the velocity dispersion of the turbulence.

In the case of the high-$\sigma$ perpendicular RMHD shock, 
the physical variables describing the postshock region in the shock rest frame are written as follows~\citep{1984ApJ...283..694K}: 
\begin{equation}
  B_2 \simeq B_1, \quad \rho_2 \gamma_2 \simeq \rho_1 \gamma_1, \quad 
\gamma_2 \simeq \sqrt{\sigma}, \quad k_B T_2 / m c^2 \simeq \gamma_1 / 8 \sqrt{\sigma},
\label{eq:RankHug}
\end{equation}
where subscripts $1$ and $2$ refer to preshock and postshock, respectively, 
$\rho$ is the mass density in the plasma comoving frame, 
$\gamma$ is the Lorentz factor in the shock front rest frame, 
$m$ and $c$ are the particle rest mass and light velocity, respectively, 
and $\sigma$ is the magnetization parameter in the preshock flow defined as: 
\begin{equation}
  \sigma = B_1^2 / 4 \pi \rho_1 c^2 \gamma_1^2
  .
\end{equation}
From the above relations, the thermal energy density in the postshock rest frame is given by 
\begin{equation}
E_{th} = \rho_1 c^2 \gamma_1^2 / 8 \sigma (\Gamma - 1), 
\end{equation}
where $\Gamma$ is the adiabatic index, 
and here we use the value in the relativistic limit, $\Gamma = 4 / 3$.
In the postshock region of the high-$\sigma$ flow, 
the dominant inertia of the plasma is due to the magnetic field energy. 
Thus, the kinetic energy density of the turbulence in the postshock rest frame can be evaluated as
\begin{equation}
  E_{turb} = [\rho_2 c^2 + B_2^2/ 4 \pi \gamma_2^2] \gamma_{turb} (\gamma_{turb} - 1) \simeq (B_2 \Delta v_{turb})^2 / 8 \pi \sigma,
\end{equation}
where $\Delta v_{turb}$ is the velocity dispersion of the turbulence. 
Since the kinetic energy density of the turbulence should be less than the thermal energy density ($E_{turb} < E_{th}$), 
the maximum velocity dispersion of turbulence induced by the shock can be evaluated as
\begin{equation}
  \Delta v_{turb} / c_s \lesssim 1.5 / \sqrt{\sigma},
  \label{eq:vmax} 
\end{equation}
where we have used $\sigma = B_1^2 / 4 \pi \rho_1 c^2 \gamma_1^2$, and the relativistic sound speed $c_s = c / \sqrt{3}$. 

According to theoretical studies of striped winds 
~\citep{2001ApJ...547..437L,2003ApJ...591..366K}, 
the magnetization parameter immediately upstream of the termination shock is estimated as $\sigma\sim100$.
In this case, Eq. (\ref{eq:vmax}) suggests that 
turbulence with $\Delta v_{turb} \lesssim 0.1\,c_s$ can be induced in the postshock flow.
Note that, 
according to the relativistic MHD simulations by Inoue et al. (2011), 
the kinetic energy of turbulence can be comparable to the released particle kinetic energy 
when the preshock density dispersion is comparable to the mean density, 
and  the velocity dispersion of turbulence depends linearly on the density dispersion. 
This suggests that 
the maximum velocity dispersion given in Eq. (5) is the case 
when the preshock density dispersion is comparable to the mean density.

In the above discussion, 
we consider the RMI as a candidate mechanism for driving turbulence in the postshock region. 
There are, however, other instabilities in the pulsar wind, 
for example, a Rayleigh-Taylor or Kruskal-Schwarzschild type instability of current sheets~\citep{2010ApJ...725L.234L} 
that may also drive turbulence in a striped wind.

\section{\label{sec:sec2}Numerical Setup}
We model the evolution of the current sheet in turbulence using 2-dimensional resistive relativistic magnetohydrodynamics (RRMHD). 
Since this calculation includes relativistically hot plasma, turbulence and magnetic dissipation, 
we need a highly accurate numerical scheme that can treat relativistic Ohmic dissipation and the evolution of magnetic field in a turbulent medium.
We use a multi-dimensional extension of the resistive RMHD scheme developed by Takamoto \& Inoue (2011). 
\footnote{
It is widely known that 
we must take into account the evolution of the electric field for satisfying causality; 
this essentially changes the mathematical properties of the governing equations, 
and makes it difficult to develop an accurate numerical scheme of RRMHD~\citep{2007MNRAS.382..995K,2009MNRAS.394.1727P,2011ApJ...735..113T}. 
An analogous problem also exists in the description of other dissipation mechanisms in the framework of relativity
~\citep{1985PhRvD..31..725H,2010PhyA..389.4580T,2011JCoPh.230.7002T}. 
}
.
This scheme uses appropriate characteristic velocities for calculating numerical fluxes, 
i.e., fast magnetosonic velocity and Alfv\'en velocity, 
so that it describes turbulent flows accurately (see Takamoto \& Inoue 2011 for detail).
We calculate the RRMHD equations in a conservative fashion, 
and the mass density, momentum, and energy are also conserved within machine round-off error. 
For the equation of state, 
we assume a relativistic ideal gas with $h = 1 + (\Gamma / (\Gamma - 1))(p_{gas} / \rho)$ where $\Gamma = 4 / 3$ and $h$ is the relativistic enthalpy. 
The resistivity, $\eta$, is assumed to be constant.
We examine the following five cases: $\eta / c^2 t_0 = 1, 2, 4, 8, 16 \times10^{-4}$ to study the effects of resistivity, 
where $t_0$ is the light crossing time for the numerical domain.

In our simulations, 
we model turbulent flow in the postshock region generated by a Poynting flux-dominated flow in the post-shock rest frame. 
Since we consider the situation of a very strong magnetic field, 
turbulence is highly sub-Alfv\'enic.
Recent studies have shown that sub-Alfv\'enic turbulence is highly anisotropic 
and that the structure of turbulent eddies is elongated along the mean magnetic field
~\citep{1995ApJ...438..763G,2001ApJ...554.1175M,2002ApJ...564..291C,2011ApJ...734...77I,2012ApJ...744...32Z}. 
In the limit of a strong magnetic field, 
turbulence can be approximated as two-dimensional (2D) in the plane perpendicular to the mean magnetic field.
Hence, as a first step in this kind of simulation, 
we perform two-dimensional simulations in the plane perpendicular to the magnetic field.
This restriction greatly reduces the computational cost 
and enables us to study the dependence of basic parameters, 
such as the resistivity and the strength of turbulence, 
on magnetic field dissipation.
The two dimensional approximation also limits magnetic energy dissipation to Ohmic dissipation, 
so that neither Sweet-Parker nor Petschek type magnetic reconnection~\citep{2000mrp..book.....B,2000mare.book.....P} can occur. 
However, 
the characteristic timescales of the relativistic tearing instability and the Sweet-Parker reconnection is given by 
$\tau_{rec} \sim (L / c_A) \sqrt{\textrm{Rm}}$, 
where $c_A$ is the Alfv\'en velocity, Rm is the magnetic Reynolds number defined as Rm $\equiv L v / \eta$, 
$L$ is the characteristic length scale, $v$ is the characteristic velocity and $\eta$ is the resistivity~\citep{2005MNRAS.358..113L}. 
On the other hand, 
the timescale of turbulence is given by eddy-turnover time $L / \Delta v$, 
where $\Delta v$ is the velocity dispersion. 
The ratio of these two timescales is 
$\tau_{turb} / \tau_{rec} \sim (L / \Delta v) / (L / c_A \sqrt{\textrm{Rm}}) \sim (c_A / \Delta v) / \sqrt{\textrm{Rm}}$, 
indicating that the effect of turbulence becomes faster than the Sweet-Parker type reconnection when $\Delta v / c_A  > 1 / \sqrt{\textrm{Rm}}$
\footnote{
If we consider the Petschek-type reconnection 
whose characteristic timescale is given by $\tau_{rec} \sim (L / c_A) \log(\mathrm{Rm})$, 
the magnetic field dissipation through reconnection might be more effective when $\Delta v / c_A < 1 / \log(\mathrm{Rm})$ 
}. 
Note that in the realistic three-dimensional case, even if the Sweet-Parker type reconnection is slow, 
the turbulence might induce ``turbulent reconnection''~\citep{1999ApJ...517..700L,2009ApJ...700...63K}. 
The 3D effect will be studied in our future paper.

For our numerical calculations, 
we prepare a square domain of $[L_x \times L_y]= [1 \times 1]$, 
and divide into the homogeneous numerical meshes of $N_x \times N_y = 1024^2$ . 
We set periodic boundary conditions for the x and y-directions.
For the initial condition, 
we consider the following magnetic field profile: 
$B_z = B_0$ for $x < 0.25 L_x$, or $x > 0.75 L_x$, 
and $B_z = - B_0$ for $0.25 L_x < x < 0.75 L_x$. 
In addition, 
we consider the relativistically hot postshock region of the Poynting flux-dominated cold preshock flow. 
According to the shock jump conditions of high-$\sigma$ flow (see Eq. [\ref{eq:RankHug}]), 
the plasma $\beta$ of the postshock region is approximately given by $\beta \sim \sigma^{-2}$ 
where $\beta$ is defined as: $\beta \equiv p_{gas} / (B^2 / 8 \pi)$. 
Thus, for a flow of $\sigma\sim 100$, 
we have $\beta\sim 10^{-4}$.
Unfortunately, 
our code having strict conservation properties cannot stably solve such a low $\beta$ plasma.
In order for the stability, 
we choose $\beta = 4$, i.e., 
we adopt a plasma with higher thermal pressure.
The bottom panel of Fig. \ref{fig:2} is the results of simulations in the case of $\beta = 0.5, 1, 4$ along with $\Delta v = 0.08 c_s$ and $k_c=3$. 
This shows that small number of plasma $\beta < 0.5$ produces essentially the same dissipation rate, 
because the rate of the dissipation does not essentially depend on the $\beta$ (see, eq. [8] below) 
\footnote{
Note that the total dissipated energy increases with decreasing $\beta$. 
This is because very small-scale corrugations develop around the current sheets in the case of very low plasma $\beta$. 
The growth of these corrugations seems analogous to the instability found by Lyubarsky \citep{2010ApJ...725L.234L}, 
and these corrugations slightly enhance the dissipation in the early phase. 
In the subsequent paper of Inoue (2012), 
it will be shown that 
the interaction between shocks and current sheets induces instability analogous to that found by Lyubarsky (2010). 
}. 
Note that the postshock state of the high-$\sigma$ flow is relativistically hot ($c_s=c/\sqrt{3}$), 
so that our choice of a larger plasma $\beta$ does not lead to an overestimation of the sound speed in the postshock region. 

In this study, 
we consider so-called ``decaying turbulence'', i.e., the turbulence is driven only initially. 
This is because the shock induced turbulence (e.g. by the RMI) is expected to be triggered only at the shock front.
We set a divergence-free initial turbulent velocity field 
whose one-dimensional power spectrum is flat 
with a cut off wave number $k_c=3/L_x$ or $5/L_x$, 
i.e., the initial turbulence is driven only at large scales. 
We study the cases of the following five initial velocity dispersions: 
$\Delta v / c_s \equiv \sqrt{\langle v^2 \rangle} / c_s
= 0.05, 0.08, 0.1, 0.3, 0.5$ 
where $c_s$ is the sound velocity. 
These parameters are motivated by the energetics of the turbulence we estimated in Sec. 2 
and the study of RMI driven turbulence by Inoue et al. (2011). 
in which the velocity dispersion of turbulence is shown to be limited to below the sound speed. 

\section{\label{sec:sec3}Results of 2D RRMHD Simulation}

\begin{figure}[h]
 \centering
  \includegraphics[width=4.2cm,clip]{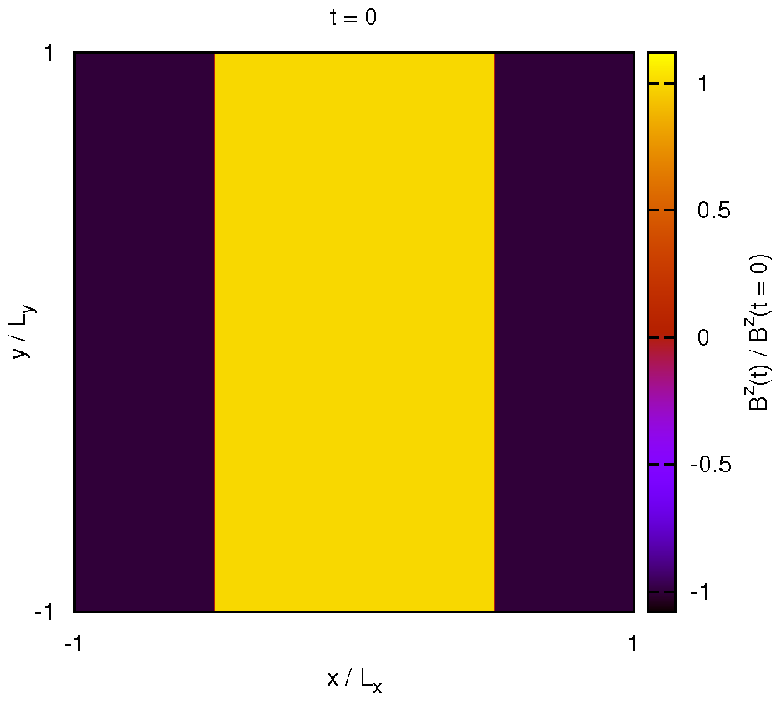}
  \includegraphics[width=4.2cm,clip]{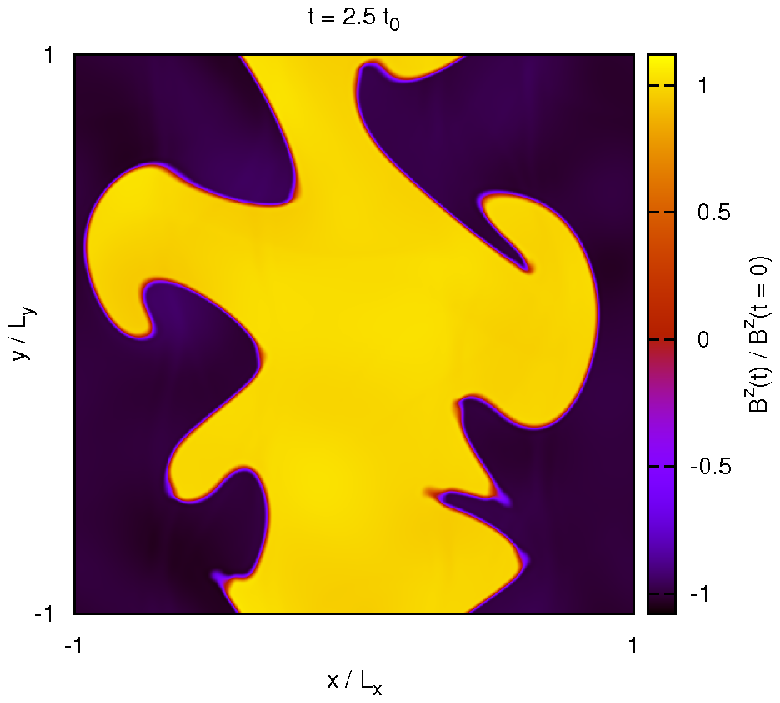}
  \includegraphics[width=4.2cm,clip]{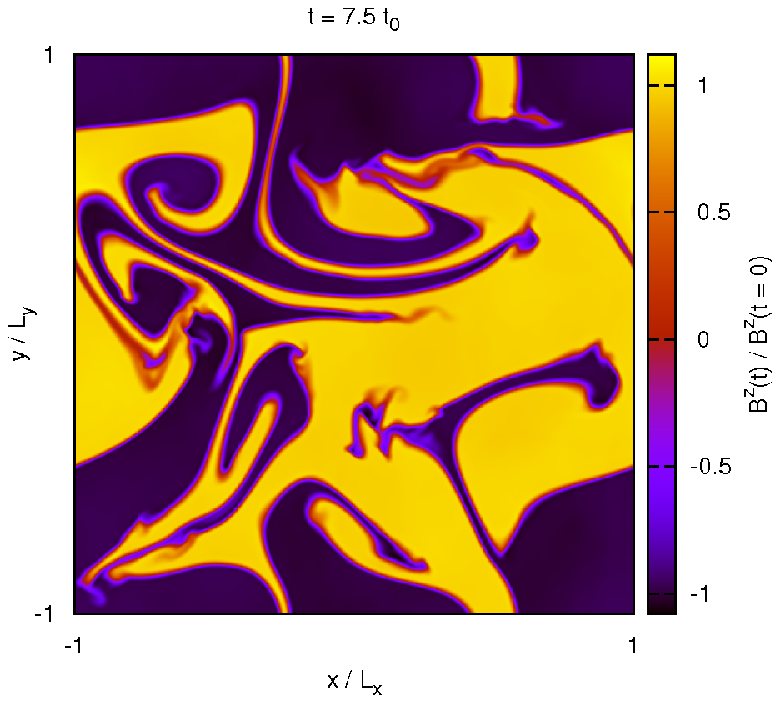}
  \includegraphics[width=4.2cm,clip]{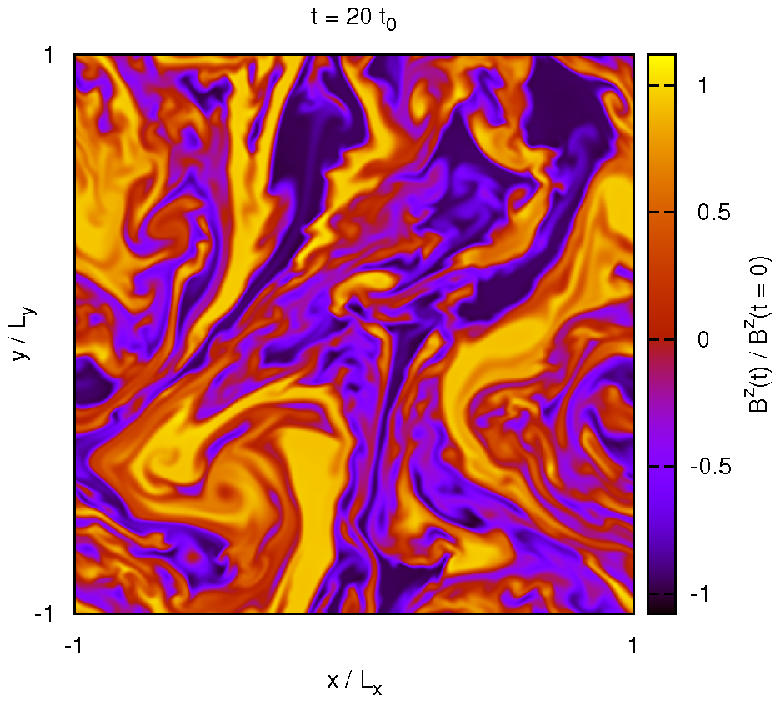}
  \caption{Snapshots of the structure of perpendicular magnetic field $B_z$ 
           at $t = 0, 2.5 t_0, 7.5 t_0, 20 t_0$ for $\Delta v = 0.1 c_s$, $\eta = 10^{-4} c^2 t_0$, $k_c = 3$, 
           where $t_0 \equiv L_x / c$. 
          }
  \label{fig:1}
\end{figure}

Fig. \ref{fig:1} shows snapshots of the structure of the perpendicular magnetic field, $B_z$, 
at $t = 0, 2.5 t_0, 7.5 t_0$ and $20 t_0$ for $\Delta v = 0.1 c_s$, $\eta = 10^{-4} c^2 t_0$, $k_c = 3$, 
where $t_0 \equiv L_x / c$.
The turbulence stretches the current sheets 
which enhances the dissipation of magnetic field. 

\begin{figure}[h]
 \centering
  \includegraphics[width=7cm,clip]{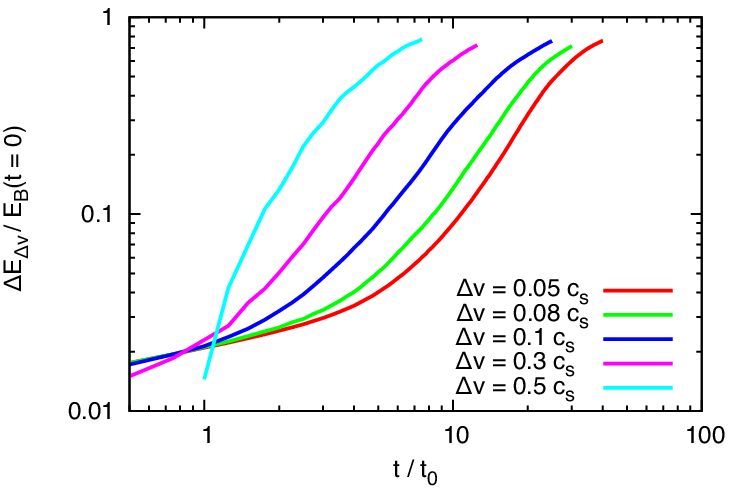}
  \includegraphics[width=7cm,clip]{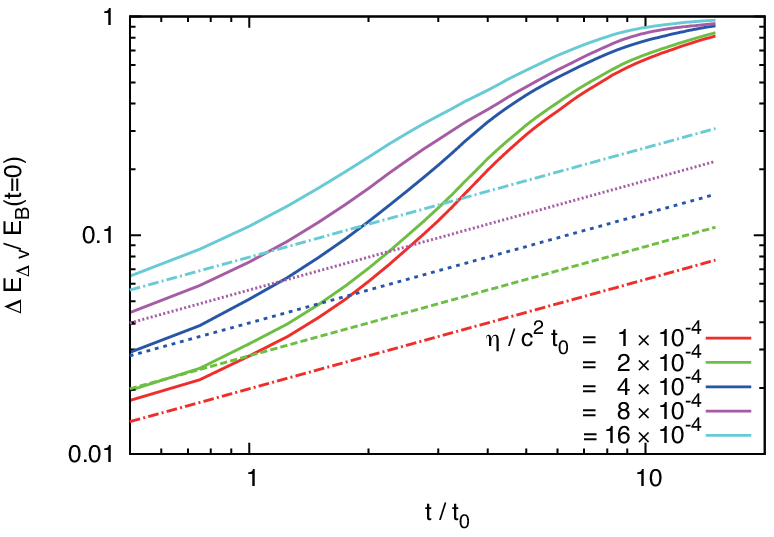}
  \includegraphics[width=7cm,clip]{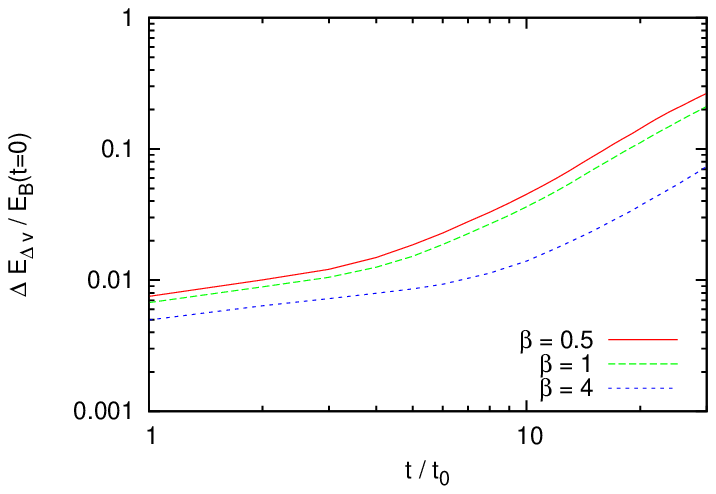}
  \caption{Temporal evolution of the total dissipated magnetic field energy. 
           Top: 
           the numerical calculations have been performed with respect to velocity dispersions 
           $\Delta v /c_s=$ 0.05, 0.08, 0.1, 0.3, and 0.5. 
           Middle: 
           the numerical calculations have been performed with respect to resistivity 
           $\eta / c^2 t_0 = 1, 2, 4, 8, 16 \times 10^{-4}$ respectively, with $\Delta v = 0.2 c_s$ and $k_c=5$. 
           Solid lines are the results with turbulence. 
           Note that $\Delta v / c_s = 0.5$ case behaves somewhat differently from others in top figure. 
           This is due to the effect of compression, 
           which amplifies magnetic field in the early stage of the evolution.
           Bottom:
           the numerical calculations have been performed with respect to 
           $\beta = p_{gas} / (B^2/ 8 \pi) = 0.5, 1, 4$ respectively, with $\Delta v = 0.08 c_s$ and $k_c=3$. 
          }
  \label{fig:2}
\end{figure}

The top panel of Fig \ref{fig:2} shows the temporal evolution of dissipated magnetic field energy with respect to various velocity dispersions.
The vertical axis is the dissipated energy compared to the initial magnetic field energy.  
In this calculation, 
we set the cut-off wave number to be $k_c = 3$.

The dissipated magnetic energy is calculated as 
$\Delta E_{B(\Delta v)} \equiv \int B_0^2 / 8 \pi dS - \int (B_z)^2 / 8 \pi dS$, 
where $B_0$ is the initial magnetic field. 
The top panel of Fig. 2 shows that 
the dissipation of magnetic field becomes faster as the velocity dispersion is increased. 
Note that the dissipation rate, however, becomes independent of the velocity dispersion after the turbulence has developed 
and most of the magnetic energy is eventually dissipated.

\begin{figure}[h]
 \centering
  \includegraphics[width=7cm,clip]{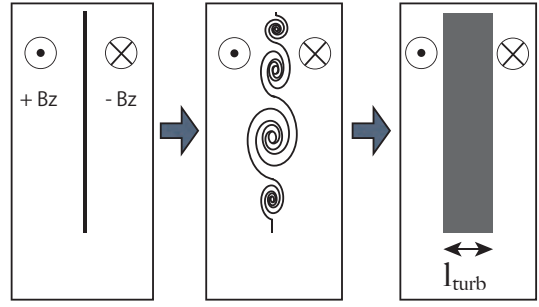}
  \caption{Schematic view of the physical mechanism of the enhancement of magnetic energy dissipation.  
           The bold lines show the current sheet. 
          }
  \label{fig:4}
\end{figure}

The middle panel of Fig. \ref{fig:2} is the dissipated magnetic energy with respect to various resistivity, $\eta$, 
with initial velocity dispersion $\Delta v = 0.2 c_s$ and $k_c=5$. 
The solid lines are the numerical results with turbulence, 
and the dashed lines are those without turbulence ($\Delta E_B\propto \eta^{1/2}\,t^{1/2}$).
In the cases without turbulence, 
the dissipated magnetic energy depends on the square root of the resistivity, as expected.
On the other hand, in the cases with turbulence, 
the dissipated magnetic energy seems to be nearly independent of the resistivity, 
in particular, after an eddy-turnover time $\sim t/t_0\gtrsim 9$.
This resistivity independent evolution of the dissipated magnetic energy can be understood as follows.
\begin{enumerate}
  \item Current sheets in turbulence are gradually stretched and coiled by eddies, 
        as shown schematically in Fig. \ref{fig:4}.

  \item The stretching of the current sheets significantly increases their surface area, 
             and the magnetic energy in an eddy dissipates quickly.

  \item This enables us to consider the effective width of the current sheet to be the eddy size, $l_{turb}$, indicating that 
             the dissipation of magnetic energy depends only on $l_{turb}$ instead of $\eta$.
\end{enumerate}

\section{Analytical Consideration}
In the following, we provide an analytical model of magnetic field dissipation in turbulence based on the above picture, 
and compare the model with our numerical simulations.
We find that our model is in good agreement with our simulations.

\begin{figure}[h]
 \centering
  \includegraphics[width=7cm,clip]{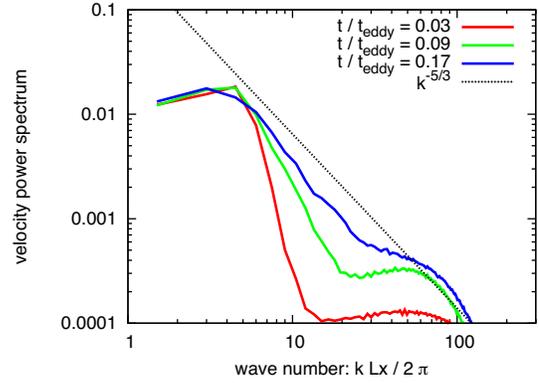}
  \caption{Temporal snapshots of one-dimensional velocity power spectra ($\int \,P_v(k) \, dk = \int v^2 d^3 x$) 
           in the case of $\Delta v = 0.2 c_s, \eta / c^2 t_0 = 1 \times 10^{-4}, k_c = 5$. 
           The spectrum quickly evolves into a Kolmogorov spectrum, $P_v(k) \propto k^{-5/3}$, 
           well within one eddy-turnover time of the initial turbulent flow ($t_{\rm eddy} \equiv L / \Delta v$). 
          }
  \label{fig:spect}
\end{figure}
\begin{figure}[h]
 \centering
  \includegraphics[width=7cm,clip]{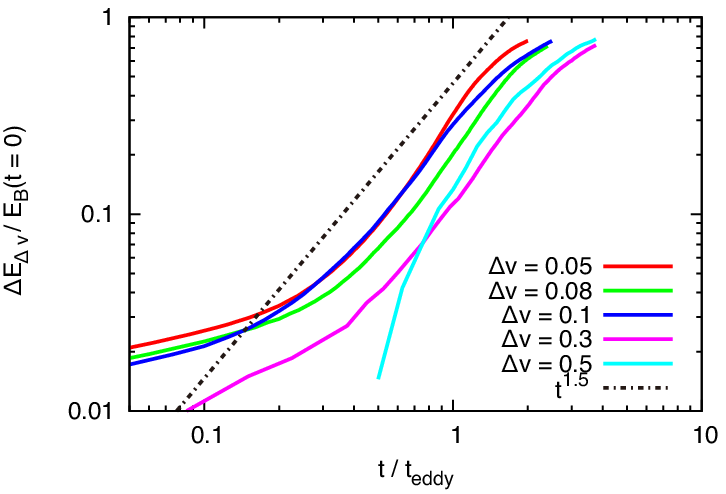}
  \includegraphics[width=7cm,clip]{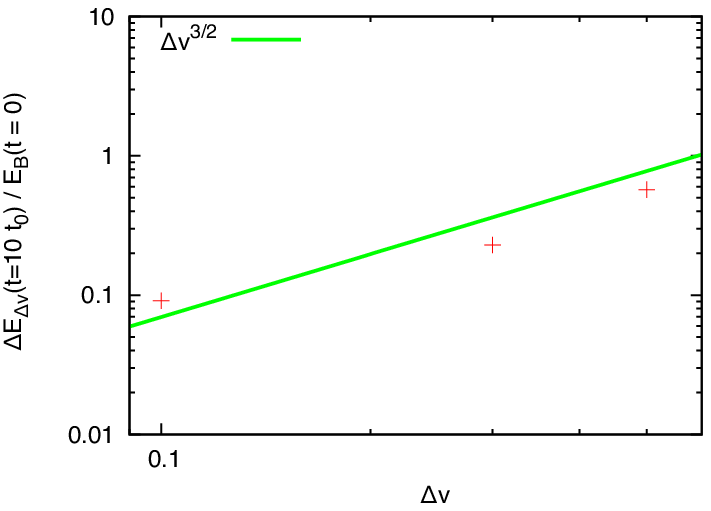}
  \caption{Top: plot of the dissipation energy, $\Delta E_{\Delta v}$, with respect to time normalized by the eddy-turnover time; $t_{eddy} \equiv L / \Delta v$.
           Bottom: dependence of the dissipation energy at $t = 10 t_0$ on the initial velocity dispersion.
          }
  \label{fig:5}
\end{figure}

The effective width of the current sheet, $l_{turb}$, is equivalent to an eddy of size 
corresponding to a turnover time equal to the elapsed time $t$ since the induction of turbulence.
In turbulence with a Kolmogorov spectrum, as realized in our simulations 
(see Fig. \ref{fig:spect}), 
it is well known that the velocity dispersion of eddies of scale $l$ is 
\begin{equation}
  v_l \sim \Delta v \left( \frac{l}{L_0} \right)^{1/3}
  \label{eq:eq1}
  ,
\end{equation}
where $L_0$ is the initial energy input scale.
Thus, the effective width of the current sheet can be written as
\begin{equation}
  l_{turb} \sim \Delta v^{3/2} t^{3/2} L_0^{-1/2}
  \label{eq:eq2}
  .
\end{equation}
The magnetic field energy in the effective current sheet is dissipated 
because current sheets are coiled by eddies; 
the dissipated magnetic energy can be evaluated as
\begin{equation}
  \Delta E_{\Delta v} \sim B_0^2 L l_{turb} \sim B_0^2 L \Delta v^{3/2} \Delta t^{3/2} / L_0^{1/2}
  \label{eq:eq3}
  ,
\end{equation}
where $L$ is the length of the current sheet. 
Eq.~(\ref{eq:eq3}) indicates that the dissipated magnetic energy, $\Delta E_{\Delta v}$, is proportional to $\Delta v^{3/2}$ and evolves as $t^{3/2}$.
Fig.~\ref{fig:5} shows the numerical results 
which confirm the above dependence. 
The top panel is a plot of the dissipated magnetic field energy $\Delta E_{\Delta v}$ with respect to time normalized by the eddy-turnover time. 
We show the results for various $\Delta v$ with fixed $\eta$ and $L_0$.
The bottom panel shows the dependence of the dissipation energy at $t = 10 t_0$ on the initial velocity dispersion.
Fig.~\ref{fig:5} clearly supports our theory.

The top panel of Fig.~\ref{fig:5}  indicates that the dissipation of magnetic field is completed in approximately a few eddy-turnover times.
The evolutionary tracks in the top panel of Fig.~\ref{fig:5} show that 
as the initial velocity dispersion becomes larger, 
the evolution normalized by the eddy-turnover time gradually becomes slower.
This is due to the effect of compression 
which amplifies magnetic field in the early stages of evolution. 
This can be also applied to the top figure of Fig. 2 
where $\Delta v / c_s = 0.5$ case behaves somewhat differently in the early phase. 

\section{\label{sec:sec4}Summary and Discussion}
In this letter, 
we have reported the acceleration of the magnetic field dissipation by decaying turbulence, 
i.e. turbulence driven only initially by an external force 
in the postshock region of a Poynting flux-dominated striped wind.
Using two-dimensional resistive relativistic MHD simulations, 
we found that owing to the stretching of current sheets, the dissipation rate of magnetic field energy is independent of resistivity 
and that most of the initial magnetic energy is dissipated within a few eddy-turnover times, 
irrespective of the initial velocity dispersion of the turbulence.
Since the interval between two successive current sheets in a striped wind is much shorter than the PWN size, 
the dissipation can be completed sufficiently close to the inner edge of the Crab PWN. 
In addition, 
Mizuno et al. (2011) have recently shown that 
the current driven instability in a PWN can reduce the $\sigma$-parameter from $0.3$ to $0.01$. 
Our dissipation mechanism can also produce a necessary setup for their mechanism of $\sigma \sim 0.3$. 
We also developed a simple analytical model of magnetic field dissipation, expressed in Eq. (\ref{eq:eq3}), 
which agrees well with the results of our simulations.

Since we assumed decaying turbulence in the postshock flow of a striped wind, 
our model can dissipate the magnetic field in the region 
where it is caught up in turbulent eddies with the current sheet.
Hence, in order to dissipate magnetic field in all regions of the striped wind, 
turbulence should be driven at least on the scale of the interval between current sheets in a striped wind.
If we consider the RMI as an origin of turbulence, 
eddies generated from inhomogeneity can be larger than the original inhomogeneity scale. 
The reason for this is as follows: 
if we consider two neighboring dense clumps 
whose distance is longer than the scale of the clumps, 
the clumps can deform a shock front 
over their separation scale, 
which can result in eddies larger than the dense clumps in the downstream flow 
(see, e.g., Inoue et al, 2012). 
This may strengthen the applicability of our mechanism. 
Furthermore, if we account for three-dimensional effects, 
we can expect the induction of turbulent reconnection~\citep{1999ApJ...517..700L,2009ApJ...700...63K}.
In that case, reconnection flows can dissipate magnetic field 
even if the initial driving scale of turbulence by the RMI is much smaller than the interval between the current sheets in the striped wind.
Since the postshock region of high-$\sigma$ flows still has high a Lorenz factor $\sim \sqrt{\sigma}$, 
a single shock does not sufficiently decelerate the flow, 
and thus another shock is necessary to slow down the wind to match the observations of the Crab PWN. 
This may not be difficult once the turbulence converts the magnetic energy into thermal energy in a few eddy-turnover time.
We will analyze this effect by performing 3D numerical simulation in our next papers.

\acknowledgments
We would like to thank referees for many fruitful comments. 
We also would like to thank Jennifer M. Stone for improving the English. 
Numerical computations were carried out in part on the Cray XT4 
at Center for Computational Astrophysics, CfCA, of National Astronomical Observatory of Japan.
Calculations were also carried out on SR16000 at YITP in Kyoto University.
This work is supported by Grant-in-aids from the Ministry of Education, Culture, Sports, Science, and Technology (MEXT) of Japan, 
No.22$\cdot$3369 and No. 23740154 (T. I.). 








\begin{thebibliography}{}
\bibitem[Biskamp (2000)]{2000mrp..book.....B} 
Biskamp, D.\ 2000, Magnetic reconnection in plasmas, Cambridge, UK: Cambridge University Press, 2000 
\bibitem[Cho et al. (2002)]{2002ApJ...564..291C} 
Cho, J., Lazarian, A., \& Vishniac, E.~T.\ 2002, \apj, 564, 291 
\bibitem[Giacalone \& Jokipii (2007)]{2007ApJ...663L..41G} 
Giacalone, J., \& Jokipii, J.~R.\ 2007, \apjl, 663, L41 
\bibitem[Giannios \& Spruit (2006)]{2006A&A...450..887G} 
Giannios, D., \& Spruit, H.~C.\ 2006, \aap, 450, 887 
\bibitem[Goldreich \& Sridhar (1995)]{1995ApJ...438..763G} 
Goldreich, P., \& Sridhar, S.\ 1995, \apj, 438, 763 
\bibitem[Hiscock \& Lindblom (1985)]{1985PhRvD..31..725H} 
Hiscock, W.~A., \& Lindblom, L.\ 1985, \prd, 31, 725 
\bibitem[Inoue et al. (2011)]{2011ApJ...734...77I} 
Inoue, T., Asano, K., \& Ioka, K.\ 2011, \apj, 734, 77 
\bibitem[Inoue et al. (2009)]{2009ApJ...695..825I} 
Inoue, T., Yamazaki, R., \& Inutsuka, S.\ 2009, \apj, 695, 825 
\bibitem[Inoue et al. (2010)]{2010ApJ...723L.108I} 
Inoue, T., Yamazaki, R., \& Inutsuka, S.\ 2010, \apjl, 723, L108
\bibitem[Inoue et al. (2012)]{2012ApJ...744...71I} 
Inoue, T., Yamazaki, R., Inutsuka, S., \& Fukui, Y.\ 2012, \apj, 744, 71 
\bibitem[Kennel \& Coroniti (1984a) ]{1984ApJ...283..694K} 
Kennel, C.~F., \& Coroniti, F.~V.\ 1984, \apj, 283, 694 
\bibitem[Kennel \& Coroniti (1984b)]{1984ApJ...283..710K} 
Kennel, C.~F., \& Coroniti, F.~V.\ 1984, \apj, 283, 710 
\bibitem[Kirk \& Skj{\ae}raasen (2003)]{2003ApJ...591..366K} 
Kirk, J.~G., \& Skj{\ae}raasen, O.\ 2003, \apj, 591, 366 
\bibitem[Komissarov et al. (2007)]{2007MNRAS.374..415K} 
Komissarov, S.~S., Barkov, M., \& Lyutikov, M.\ 2007, \mnras, 374, 415
\bibitem[Komissarov (2007)]{2007MNRAS.382..995K} 
Komissarov, S.~S.\ 2007, \mnras, 382, 995
\bibitem[Kowal et al. (2009)]{2009ApJ...700...63K} 
Kowal, G., Lazarian, A., Vishniac, E.~T., \& Otmianowska-Mazur, K.\ 2009, \apj, 700, 63 
\bibitem[Lazarian \& Vishniac (1999)]{1999ApJ...517..700L} 
Lazarian, A., \& Vishniac, E.~T.\ 1999, \apj, 517, 700 
\bibitem[Lyubarsky (2005)]{2005MNRAS.358..113L} 
Lyubarsky, Y.~E.\ 2005, \mnras, 358, 113 
\bibitem[Lyubarsky (2010)]{2010ApJ...725L.234L} 
Lyubarsky, Y.\ 2010, \apjl, 725, L234 
\bibitem[Lyubarsky \& Kirk (2001)]{2001ApJ...547..437L} 
Lyubarsky, Y., \& Kirk, J.~G.\ 2001, \apj, 547, 437 
\bibitem[Lyutikov \& Blandford (2003)]{2003astro.ph.12347L} 
Lyutikov, M., \& Blandford, R.\ 2003, arXiv:astro-ph/0312347
\bibitem[Maron \& Goldreich (2001)]{2001ApJ...554.1175M} 
Maron, J., \& Goldreich, P.\ 2001, \apj, 554, 1175 
\bibitem[Meier (2003)]{2003NewAR..47..667M} 
Meier, D.~L.\ 2003, New A Rev., 47, 667 
\bibitem[Meszaros \& Rees (1997)]{1997ApJ...482L..29M} 
Meszaros, P., \& Rees, M.~J.\ 1997, \apjl, 482, L29
\bibitem[Michel (1971)]{1971CoASP...3...80M} 
Michel, F.~C.\ 1971, Comments on Astrophysics and Space Physics, 3, 80 
\bibitem[Mizuno et al. (2011)]{2011ApJ...728...90M} 
Mizuno, Y., Lyubarsky, Y., Nishikawa, K.-I., \& Hardee, P.~E.\ 2011, \apj, 728, 90 
\bibitem[Palenzuela et al. (2009)]{2009MNRAS.394.1727P} 
Palenzuela, C., Lehner, L., Reula, O., \& Rezzolla, L.\ 2009, \mnras, 394, 1727
\bibitem[P{\'e}tri \& Lyubarsky (2007)]{2007A&A...473..683P} 
P{\'e}tri, J., \& Lyubarsky, Y.\ 2007, \aap, 473, 683
\bibitem[Priest \& Forbes (2000)]{2000mare.book.....P} 
Priest, E., \& Forbes, T.\ 2000, Magnetic Reconnection, by Eric Priest and Terry Forbes, pp.~612.~ISBN 0521481791.~Cambridge, 
UK: Cambridge University Press, June 2000.,  
\bibitem[Sironi \& Goodman (2007)]{2007ApJ...671.1858S} 
Sironi, L., \& Goodman, J.\ 2007, \apj, 671, 1858 
\bibitem[Sironi \& Spitkovsky (2011)]{2011ApJ...741...39S} 
Sironi, L., \& Spitkovsky, A.\ 2011, \apj, 741, 39 
\bibitem[Sturrock et al. (1984)]{1984SoPh...94..341S} 
Sturrock, P.~A., Kaufman, P., Moore, R.~L., \& Smith, D.~F.\ 1984, \solphys, 94, 341
\bibitem[Takamoto \& Inoue (2011)]{2011ApJ...735..113T} 
Takamoto, M., \& Inoue, T.\ 2011, \apj, 735, 113
\bibitem[Takamoto \& Inutsuka (2010)]{2010PhyA..389.4580T} 
Takamoto, M., \& Inutsuka, S.\ 2010, Physica A Statistical Mechanics and its Applications, 389, 4580 
\bibitem[Takamoto \& Inutsuka (2011)]{2011JCoPh.230.7002T} 
Takamoto, M., \& Inutsuka, S.\ 2011, Journal of Computational Physics, 230, 7002 
\bibitem[Zrake \& MacFadyen (2012)]{2012ApJ...744...32Z} 
Zrake, J., \& MacFadyen, A.~I.\ 2012, \apj, 744, 32 
\end{thebibliography}

\end{document}